\documentclass[a4paper]{jpconf}
\usepackage{graphicx}
\usepackage{subfig}
\usepackage{sidecap}
\usepackage{hyperref}
\begin{document}
\title{The optical simulation model of the DarkSide-20k Veto detector}

\author{Cenk T\"urko\u{g}lu and Sarthak Choudhary for the DarkSide collaboration}

\address{AstroCeNT, Nicolaus Copernicus Astronomical Center, Polish Academy of Sciences, Rektorska 4, 00-614 Warsaw, Poland}

\ead{cturkoglu@camk.edu.pl}

\begin{abstract}
DarkSide-20k is a rare-event search experiment aimed at finding signals of dark matter particles. It is a dual-phase detector that registers ionisation and scintillation signals originating from the particles interacting with the liquid argon detector medium. It is enclosed in a single-phase liquid argon neutron veto tank, equipped with Gd-loaded panels for capturing neutrons. Since vetoing and particle identification are carried out using the light signal, it is crucial to maximise the light yield. Light collection efficiency depends on optical properties of the detector and particularly for the veto detector, which has a photosensor coverage of the order of a per cent, the reflectivity of the walls has a big impact. To quantify the amount of collected light, a comprehensive Geant4 simulation is performed, which uses optical characterisation data. In this work, a detailed description of the optics model for the veto of the experiment will be discussed.
\end{abstract}

\section{Introduction}
DarkSide-20k (DS-20k) is a dual-phase rare-event search experiment dedicated to finding the signals of dark matter particles. To find this signal, DS-20k employs liquid argon (LAr) as the detector medium, and uses ionisation and scintillation-light signals to detect particle interactions. LAr is used in the inner time projection chamber (TPC) ($\sim$~50 tons) instrumented with SiPMs which is enclosed in an active LAr veto, also enclosed in a passive LAr shield tank \cite{ds20k}. The veto of the DS-20k is crucial for detecting neutrons that can mimic dark matter induced signals. Neutrons are moderated by collisions in the acrylic.  Gd in the acrylic ensures the emission of multiple high energy (with a threshold of $\mathrm{^{39}Ar}$ endpoint to define hits) $\gamma$-rays after the neutron capture.  For both scintillation and ionisation channels particle detection is carried out via light signals, light collection becomes an essential part for both the TPC and the veto. 

\begin{figure}
\begin{minipage}{.5\linewidth}
\centering
\subfloat[]{\label{ds20k}\includegraphics[scale=.3]{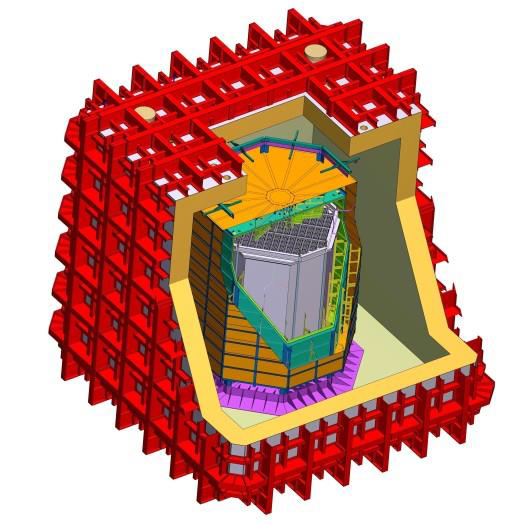}}
\end{minipage}%
\begin{minipage}{.5\linewidth}
\centering
\subfloat[]{\label{layers}\includegraphics[scale=.10]{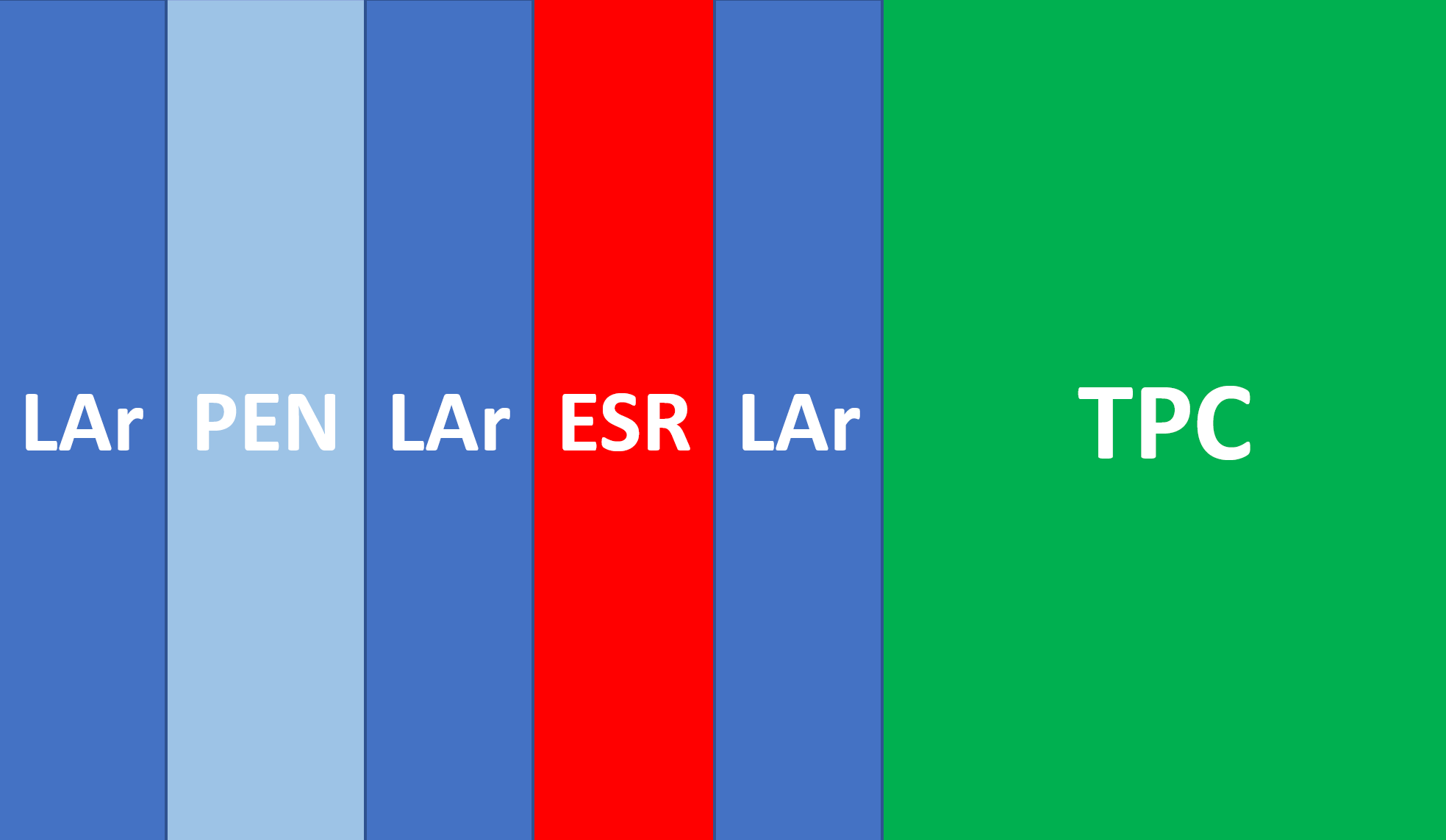}}
\end{minipage}\par\medskip
\caption{(\ref{ds20k}) DarkSide-20k inside the ProtoDUNE cryostat \cite{ds20k} and (\ref{layers}) layers of the veto made up of LAr, ESR and PEN}
\label{fig:design}
\end{figure}

\section{Simulation}
In order to test the performance of the DS-20k veto and optimise it before finalising the design, a Geant4 simulation code "g4ds" \cite{g4ds} was developed. The simulation code encapsulates a detailed geometry of the DS-20k detector and the physics processes that are expected to occur in the experiment. The simulation includes the geometry to the TPC, of the veto, and of the shield tank. One of the most important aspects of the veto simulation is the optics since the light collection efficiency can change dramatically depending on the materials used and the design. For this reason, a detailed optical model was implemented in the simulation code. \\

SiPMs are not sensitive to vacuum ultraviolet (VUV) scintillation light emitted from particle interactions with LAr. Hence, wavelength shifter (WLS) materials that can convert the wavelength of the incoming VUV photons into optical photons are used both inside the TPC and in the veto. While in the TPC, a vacuum evaporated tetraphenyl-butadiene (TPB) coating is used as the WLS of choice, in the veto, polyethylene naphthalate (PEN) films are used. Both of these materials have advantages with respect to each other~\cite{reflectivities}. Since PEN is in the form of a thin film, various optical parameters should be taken into account and used in the simulations. In addition to the WLSs, design and materials used in the setup have parameters that can affect the light collection.

\subsection{WLS Properties} \label{WLS}
Since PEN is available as a large format film, it is much more scalable compared to TPB, which makes it attractive for DS-20k and future liquid argon based detectors. There various optical properties of PEN which are crucial to the light collection in the veto. \\

One of the most significant parameters for WLSs is the wavelength shifting efficiency. In the beginning of the design process for the DS-20k, TPB was planned to be used as the WLS for both the veto and the TPC. However, considering the advantages of using PEN, as well as the latest results from one-to-one comparison of TPB and PEN in cryogenic conditions \cite{reflectivities} lead to the decision of using PEN as the WLS for the veto. In \cite{reflectivities}, the relative wavelength shifting efficiency (WLSE) of PEN compared to that of TPB was reported to be 47.2~$\%$. \\

In the simulation absorption length for PEN was given separate values for VUV and visible regimes. The two being separated roughly at 150~nm. The same scheme was followed for Rayleigh scattering length as well. Absorption length for the visible part of the spectrum, $\sim$1~cm~\cite{absorptionlength}. For Rayleigh scattering length, 150~$\mathrm{\mu m}$ is assumed for the visible part of the spectrum. Refractive indices for PEN both for the visible and the UV parts are taken as 1.75. Emission spectrum of PEN, another parameter taken into account, can be seen on Figure~\ref{main:a}. Re-emission time used in the simulations was observed to be 20 ns~\cite{emissiontime} for PEN. 

\subsection{Properties of the Veto Materials} \label{materials}
In addition to the parameters related to the WLSs, optical properties of other materials in the veto are also essential in the light collection. One of these properties is the absorption length of LAr, which is assumed 100~m for both visible and UV part of the spectrum. \\

\begin{figure}
\begin{minipage}{.5\linewidth}
\centering
\subfloat[]{\label{main:a}\includegraphics[scale=.14]{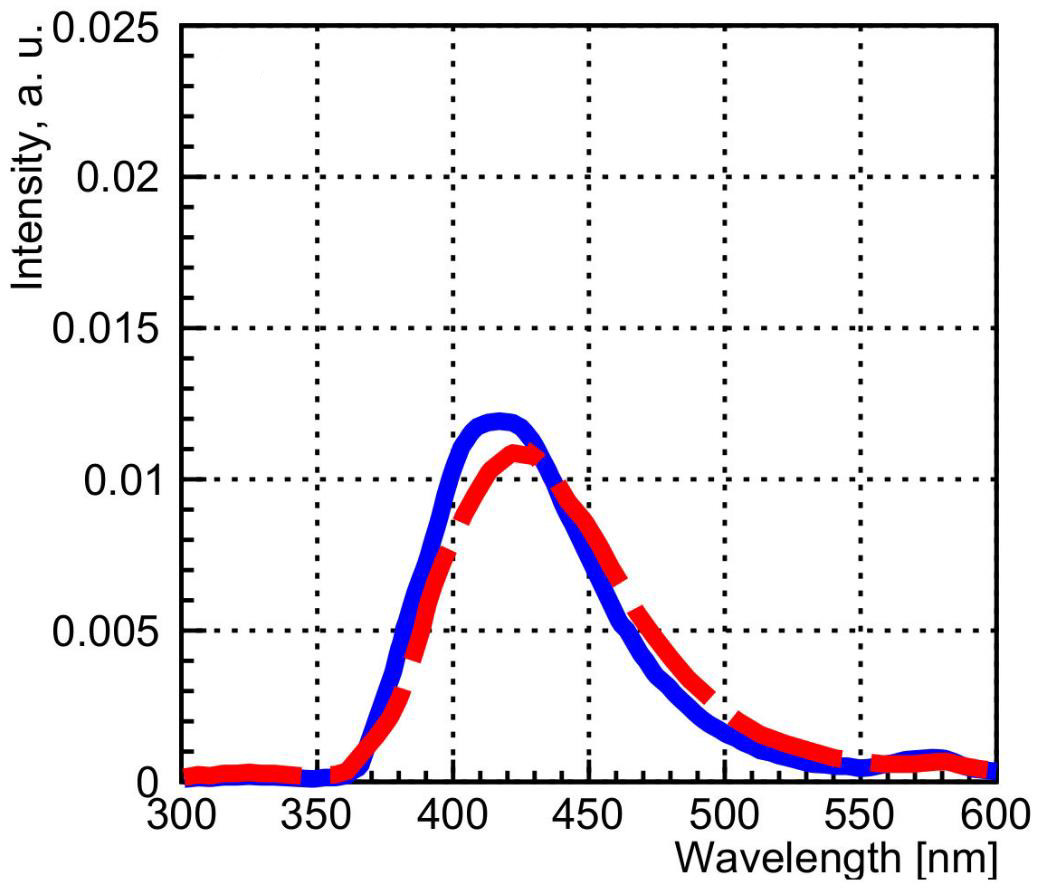}}
\end{minipage}%
\begin{minipage}{.5\linewidth}
\centering
\subfloat[]{\label{main:b}\includegraphics[scale=.15]{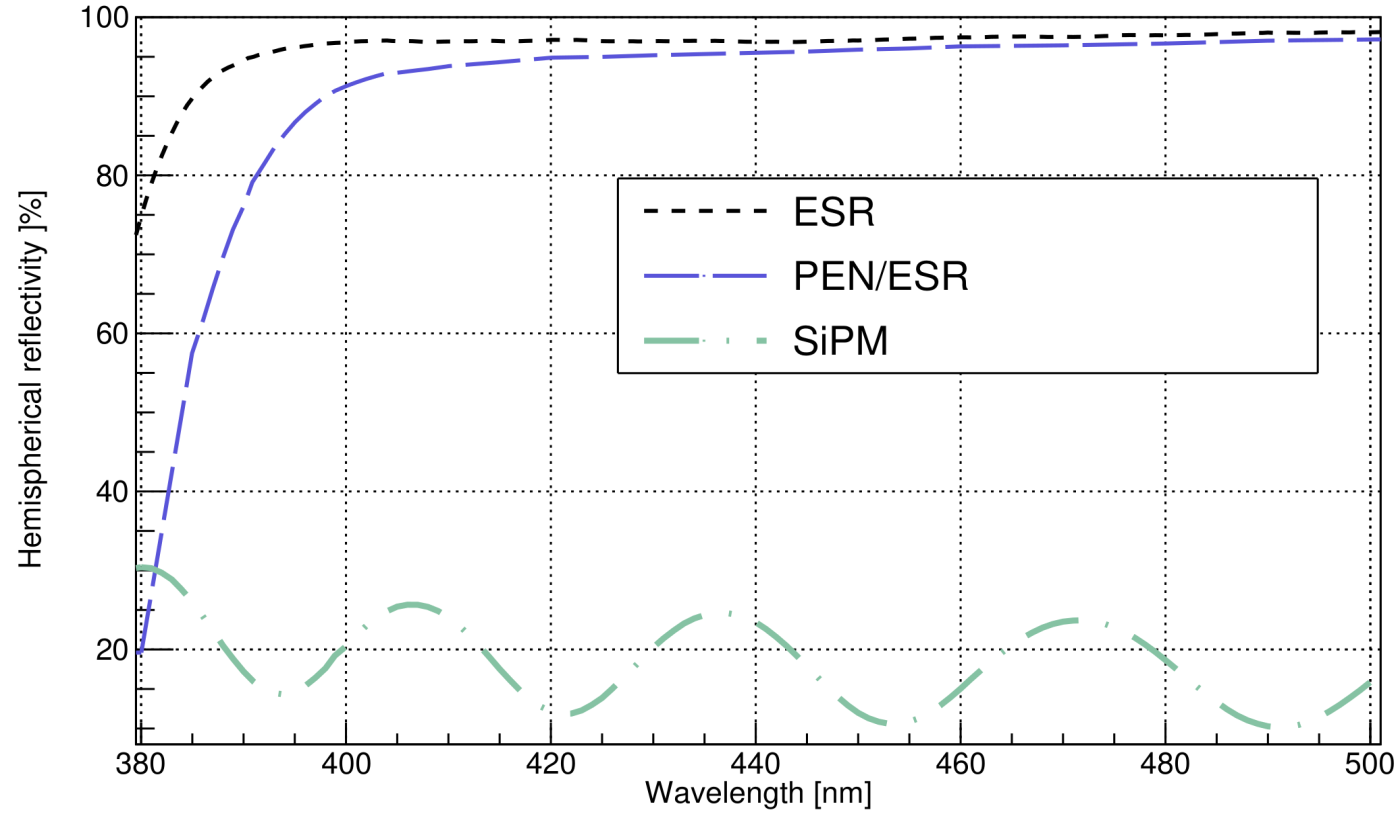}}
\end{minipage}\par\medskip
\caption{Spectra of various optical parameters used in the simulation: (\ref{main:a}) emission spectra of PEN at room temperature (red) and at cryogenic temperature (blue)~\cite{emission} and (\ref{main:b}) reflectivity percentages of ESR (black), PEN air-coupled to ESR (blue) and SiPM (green)~\cite{reflectivities}}
\label{fig:parameters}
\end{figure}

Reflectivities of other parts of the veto are crucial as even a minuscule change can change the light yield (LY). The walls of the veto detector are covered with a thin film called Enhanced Specular Reflector (ESR), which enhances the light collection of the SiPMs with its $~98~\%$ reflectivity. Reflectivity curve of ESR is shown on Figure \ref{main:b} in black dashed line. Another material with reflectivity information used in the simulation is the SiPMs. The reflectivity of the SiPM surface indirectly affects the amount of photons detected (see green dashed line on Figure~\ref{main:b}), as well as the photon detection efficiency (PDE) of the SiPM \cite{pde}.

\section{Results and Conclusions}
After inserting the parameters discussed in sections~\ref{WLS} and \ref{materials}, simulations were performed. Individual effects of these parameters were investigated in order to understand the strength of their effects on the LY. To understand whether the final results are correct or not, a comparison with a simple analytical model was made \cite{analytical}:
\begin{equation} \label{analytical_eq}
\mathrm{LY = 40 \: [ph/keV] \cdot PDE \: [pe/ph] \cdot WLSE \cdot \frac{F_{sens} \cdot FF \cdot (1 - R_{sens})}{1 - (F_{sens} \cdot R_{sens} + (1 - F_{sens}) \cdot R_{wall})}}
\end{equation}
This model takes into account the wall reflectivity ($\mathrm{R_{wall}}$), the SiPM coverage fraction ($\mathrm{F_{sens}}$), the SiPM reflectivity ($\mathrm{R_{sens}}$), the SiPM PDE (PDE), fill factor (FF) and the WLSE. \\

\begin{figure}[h]
\centering
\includegraphics[width=0.6\textwidth]{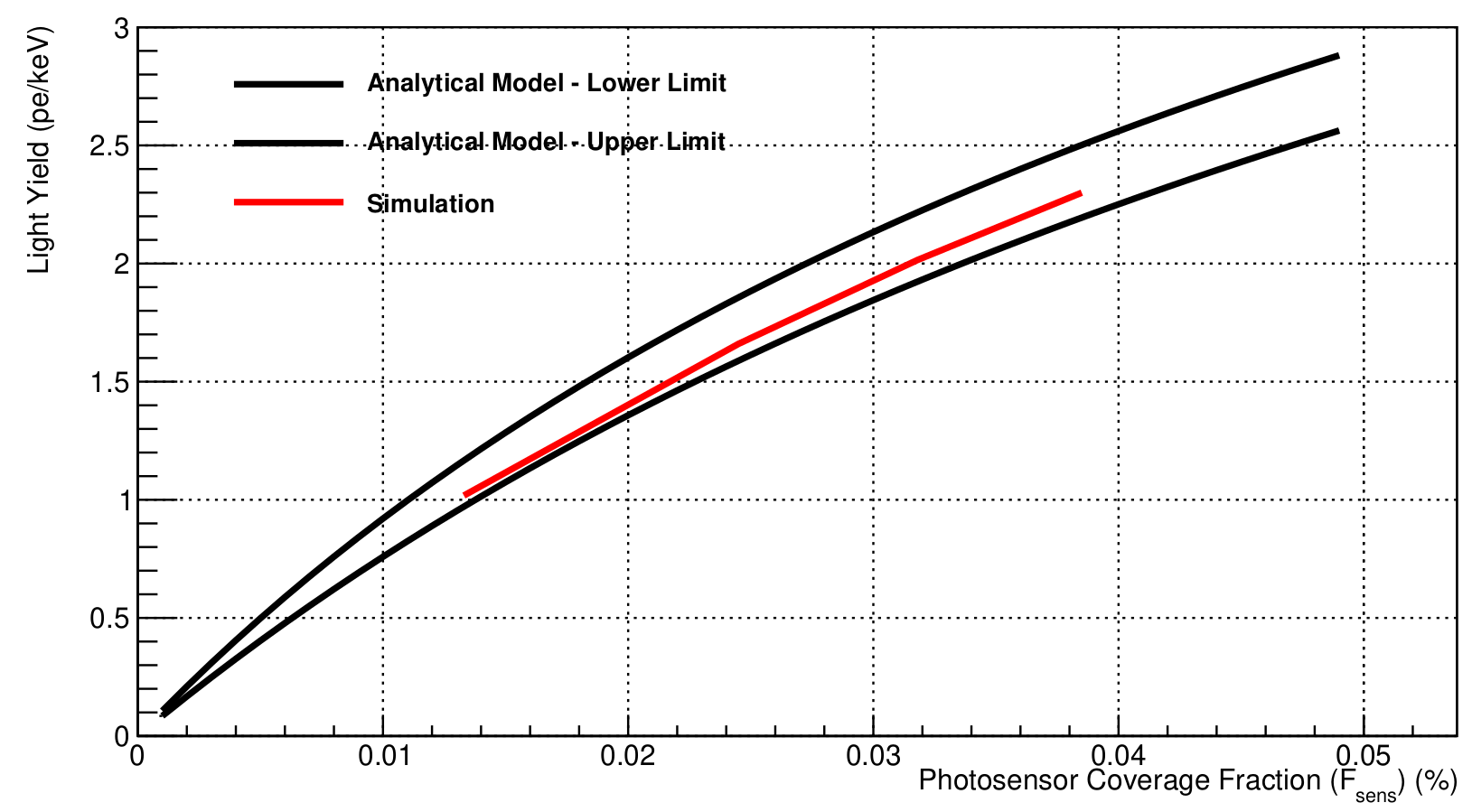}
\caption{Light yield output of the veto simulations (red) with respect to different configurations of the SiPM coverage fraction compared to the analytical model (black) according
to~\cite{analytical}}.
\label{analytical}
\end{figure}

The Geant4 simulation was validated against the analytical model, and the results can be seen on Figure \ref{analytical}. The graph shows the LY in units of photoelectrons (pe) per energy (keV), and it is drawn with respect to the photosensor (SiPM, in the case of DS-20k) coverage fraction, which is the trend future dark matter direct detection experiments are working towards. This comparison confirmed the suitability of PEN as WLS to be used in the DS-20k veto, despite its lower WLSE compared to that of TPB. 

\ack
The DarkSide Collaboration would like to thank LNGS and its staff for invaluable technical and logistical support. This report is based upon work supported by the U. S. National Science Foundation (NSF) (Grants No. PHY-0919363, No. PHY-1004054, No. PHY-1004072, No. PHY-1242585, No. PHY-1314483, No. PHY- 1314507, associated collaborative grants, No. PHY-1211308, No. PHY-1314501, No. PHY-1455351 and No. PHY-1606912, as well as Major Research Instrumentation Grant No. MRI-1429544), the Italian Istituto Nazionale di Fisica Nucleare (Grants from Italian Ministero dell’Istruzione, Università, e Ricerca Progetto Premiale 2013 and Commissione Scientific Nazionale II), the Natural Sciences and Engineering Research Council of Canada, SNOLAB, and the Arthur B. McDonald Canadian Astroparticle Physics Research Institute. We acknowledge the financial support by LabEx UnivEarthS (ANR-10-LABX-0023 and ANR-18-IDEX-0001), the São Paulo Research Foundation (Grant FAPESP-2017/26238-4), and the Russian Science Foundation Grant No. 16-12-10369. The authors were also supported by the “Unidad de Excelencia María de Maeztu: CIEMAT - Física de partículas” (Grant MDM2015-0509), the Polish National Science Centre (Grant No. UMO-2019/33/B/ST2/02884), the Foundation for Polish Science (Grant No. TEAM/2016-2/17), the International Research Agenda Programme AstroCeNT (Grant No. MAB/2018/7) funded by the Foundation for Polish Science from the European Regional Development Fund, the European Union’s Horizon~2020 research and innovation programme (DarkWave, Grant No. 962480), the Science and Technology Facilities Council, part of the United Kingdom Research and Innovation, and The Royal Society (United Kingdom). I.F.M.A is supported in part by Conselho Nacional de Desenvolvimento Científico e Tecnológico (CNPq). We also wish to acknowledge the support from Pacific Northwest National Laboratory, which is operated by Battelle for the U.S. Department of Energy under Contract No. DE-AC05-76RL01830.

\end{document}